\documentclass{raa}            % referee version: for submission

\usepackage{graphicx,times}             %for PS/EPS graphics inclusion, new

\begin{document}

   \title{A magnetically driven origin for the low luminosity GRB 170817A associated with GW170817%\,$^*$
%\footnotetext{$*$ Supported by the National Natural Science Foundation of China.}
}
%   \subtitle{I. Place Your Subtitle Here}

   \volnopage{Vol.0 (20xx) No.0, 000--000}      %%preserved for Editor. DOn't remove!
   \setcounter{page}{1}          %%starting page, preserved for Editor. DOn't remove!

   \author{H. Tong
      \inst{1}
   \and C. Yu
      \inst{2}
   \and L. Huang
      \inst{3}
   }

   \institute{School of Physics and Electronic Engineering, Guangzhou University, 
   Guangzhou 510006, China; {\it htong\_2005@163.com}\\
        \and
             School of Physics and Astronomy, Sun Yat-sen University, Zhuhai 519082, China; 
             {\it yucong@mail.sysu.edu.cn}\\
        \and
             Key Laboratory for Research in Galaxies and Cosmology, Shanghai Astronomical Observatory, 
             Chinese Academy of Sciences, Shanghai 200030, China
   }

   \date{Received~~2009 month day; accepted~~2009~~month day}

\abstract{ The gamma-ray burst GRB 170817A associated with GW170817 is subluminous and subenergetic compared with
other typical short GRBs. It may be due to a relativistic jet viewed off-axis, or a structured jet, or cocoon emission.
Giant flares from magnetars may possibly be ruled out. However, the luminosity and energetics of GRB 170817A is
coincident with that of magnetar giant flares. After the coalescence of the binary neutron star, a hypermassive neutron
star may be formed. The hypermassive neutron star may have magnetar-strength magnetic field. 
During the collapse of the hypermassive neutron star, the magnetic field energy will also be released. This giant-flare-like event may explain the the luminosity and energetics of GRB 170817A. Bursts with similar luminosity and energetics
are expected in future neutron star-neutron star or neutron star-black hole mergers.
\keywords{stars: magnetar --- stars: neutron --- gamma-ray burst: individual (GRB 170817A) --- gravitational waves}
}

   \authorrunning{Tong, Yu  \& Huang}            %author_head in even pages
   \titlerunning{Magnetically driven origin for GRB 170817A}  % title_head in odd pages

   \maketitle
%% The author head (on even pages) and the title head (on odd pages) will be
%% automatically extracted from \author{} and \title{}. Whenever the title is too long,
%% you will be asked to supply a shorter one by inserting either \authorrunning{} or
%% \titlerunning{} before \maketitle. Anyway, you can specify your own heads.
%%
%%
%% Note: In the following text body of your manuscript, please note several differences from
%%       other major journals:
%% (1) \subsection{Please Capitalize the First Letter of Each Notional Word in Subsection Title}
%% (2) Please Capitalize the First Letter of Each Notional Word in all tables' captions

%
%________________________________________________ sections below
%
\section{Introduction}           %% first-level sections will be auto-capitalized

GW170817 is the gravitational wave event  of a binary neutron star inspiral (Abbott et al. 2017a). This event also has
multi-wavelength electromagnetic counterparts (Coulter et al. 2017; Abbott et al. 2017b): a possible short gamma-ray burst (GRB), GRB 170817A (Abbott et al. 2017c; Goldstein et al. 2017; Savchenko et al. 2017);
ultraviolet/optical/infrared emissions from a kilonova (Villar et al. 2017 and references
therein); delayed X-ray and radio emission, which may be the afterglow (Troja et al. 2017; Hallinan et al. 2017).

Detailed analysis show that GRB 170817A is subluminous and subenergetic compared with other cosmological
short gamma-ray bursts (Abbott et al. 2017c; Goldstein et al. 2017; Savchenko et al. 2017; He et al. 2017; Fong et al. 2017).
Its isotropic energy and luminosity are:
$E_{\rm iso} =3.1\times 10^{46} \ \rm erg $, and $L_{\rm iso} =1.6\times 10^{47} \ \rm erg \ s^{-1}$, respectively, in the
$1\ \rm keV$-$10 \ \rm MeV$ range. It is 2 to 6 orders of magnitude less energetic than other short GRBs (Abbott et al. 2017c).
The physics for the subluminous of this GRB may be: (1) a relativistic jet viewed off-axis, (2) a structured jet, or (3) cocoon emission
(Abbott et al. 2017c; Murguia-Berthier et al. 2017; Kasliwal et al. 2017). For an off-axis jet, later deceleration may explain the delayed X-ray and radio afterglow emission (Troja et al. 2017; Hallinan et al. 2017).
However, in order to see the prompt emission of the jet, a fine tuning
of the line of sight may be required. The solid angle that we may see this GRB is very small (Abbott et al. 2017c). A slightly off-axis jet (e.g., jet
opening angle of $25^{\circ}$ and viewing angle of $30^\circ$) may be ruled out by the radio observations (Hallinan et al. 2017).
While, for a widely off-axis jet (e.g., jet opening angle $10^\circ$ and viewing angle $30^\circ$), it can explain the X-ray and radio
afterglow emissions. However, in the widely off-axis jet case, an independent mechanism for the prompt short GRB is required (Hallinan et al. 2017).
Furthermore, the energetics and luminosity of GRB 170817A is similar to the giant flare seen in a magnetar (Hurly et al. 2005; Palmer et al. 2005; Mereghetti 2008). This coincidence may be accidental. If not, it may suggest that GRB 170817A and magnetar giant flares share similar physical processes. The later probability is explored in the following.

The association of GRB 170817A with a binary neutron star merger event, soft spectrum, and lack of tail emission may argue against a magnetar giant flare origin (Abbott et al. 2017c; Goldstein et al. 2017). However, the remnant of the binary neutron star merger may be a hypermassive neutron star (Abbott et al. 2017c; Muiguia-Berthier et al. 2017). The neutron star may have magnetar strength magnetic field due to interactions between convection and differential rotation during the formation process. Subsequent collapse of the hypermassive neutron star to a black hole will also result in the release of the neutron star's magnetic energy. This giant-flare-like event may be responsible for the subluminous GRB 170817A, especially its coincidence with magnetar giant flare energetics. 

\section{Description of the scenario}

The binary neutron merger may result in a remnant with mass $2.7$-$2.8 \ \rm M_{\odot}$ (Abbott et al. 2017a). This mass lies in the hypermassive range for many neutron star equation of state (Abbott et al. 2017c). The presence of a blue kilonova may also indicate
the presence of a hypermassive neutron star (Murguia-Berthier et al. 2017). A hypermassive neutron star has mass larger than the maximum mass of a uniformly rotating neutron star. It is supported by the differential rotation. Subsequent dissipation of the differential rotation will result in a collapse of the hypermassive neutron star, to a black hole (Baumgarte et al. 2000; Hotokezaka et al. 2013). The magnetic braking or viscous dissipation timescale is about $100\ \rm ms$ (Hotokezaka et al. 2013).  However, for a nascent hypermassive neutron star, collapse can happen only when the its thermal energy is carried away by neutrinos (Sekiguchi et al. 2011).  The typical neutrino cooling timescale is order of seconds. This may corresponds to the $1.7 \ \rm s$ delay of GRB 170817A and the merger time of GW170817.
The footpoint of magnetic field lines are initially anchored to the neutron star crust. After the collapse
of the hypermassive neutron star to a black hole, there are no solid crust to be anchored for the field lines.
The magnetic field may be ejected. Reconnection may occur during this process, similar to solar coronal mass ejection and magnetar giant flares (Lyutikov 2006; Elenbaas et al. 2016).

According to Hotokezaka et al. 2013 (Figure 2 there and references therein), the nascent hot neutron star may be $0.1\ \rm M_{\odot}$  heavier than the maximum mass of a cold supramassive neutron star. The maximum mass of a cold supramassive neutron star is about $1.2$ times the maximum mass of  a cold nonrotating neutron star. Assuming a remnant mass of $2.75 \ \rm M_{\odot}$ (Abbott et al. 2017a), the above scenario results in a maximum mass of a cold non-rotating neutron star about $2.2 \ \rm M_{\odot}$. This rough estimation is consistent with results of more detailed analysis (Margalit \& Metzger 2017; Shibata et al. 2017; Rezzolla et al. 2018; Ruiz et al. 2018).

Prior study of the double pulsar system shows that the binary neutron star may be made up of a normal neutron star and a recycled one (Lyne et al. 2004).  For a merger time scale of $10\ \rm Gyr$ (Blanchard et al. 2017), the magnetic field of the two neutron stars may have decayed significantly. During the birth of the hypermassive neutron star, it may acquire rapid rotation and strong magnetic field (Zrake \& MacFadyen 2013; Giacomazzo \& Perna 2013; Ciolfi et al. 2017). The turbulent dynamo process for normal magnetars may also take place in the case of  nascent hypermassive neutron stars (Duncan \& Thompson 1992). Its magnetic field can be as high as the magentar magnetic field, e.g. up to $10^{15}$-$10^{16} \ \rm G$.  For a volume about the cube of the neutron star radius $\Delta V \sim 10^{18} \ \rm cm^3$, the stored magnetic energy is about $E_{\rm mag} \sim B^2/(8\pi) \Delta V \sim 4\times 10^{46}$-$4\times 10^{48} \ \rm erg$.
This energy is enough to power the soft gamma-ray emissions of GRB 170817A (Abbott et al. 2017c; Goldstein et al. 2017). The spike of
magnetar giant flares lasts about $0.5 \ \rm s$ (Hurley et al. 2005; Palmer et al. 2005; Mereghetti 2008). The magnetic field may be dragged by the expanding ejecta and lead to formation of current sheet (Lyutikov 2006; Yu \& Huang 2013).
During the escape of the ejecta, the magnetic dissipation inside the current sheet would give rise to flares
and the duration timescale is determined by the escape velocity of the expanding ejecta, which depends closely on the magnetic reconnection inflow velocity. Our detailed calculation shows that the inflow Mach number $M_A=V_{\rm inflow}/V_A \sim V_{\rm inflow}/c$ less than  $10^{-3}$ can well reproduce the observed flare duration timescale (Yu et al. in prep.).  This may also explain why the initial pulse of GRB 170817A also lasts about $0.5 \ \rm s$ (Goldstein et al. 2017; Savchenko et al. 2017). Then the luminosity of the initial pulse should be around $10^{47} \ \rm erg \ s^{-1}$. Therefore, a giant-flare-like origin can explain the subluminous and subenergetic GRB 170817A, associated with GW170817. 

\begin{figure}
\centering
%%%/PlanetStruct/TurbSuperEarth/opacity_indx_check/work_alpha1_beta1/turb0.0/kappa1e-3
%%%PTM0010.dat , PTM0100.dat , PTM0200.dat , script in foursub.py
\includegraphics[scale=0.75]{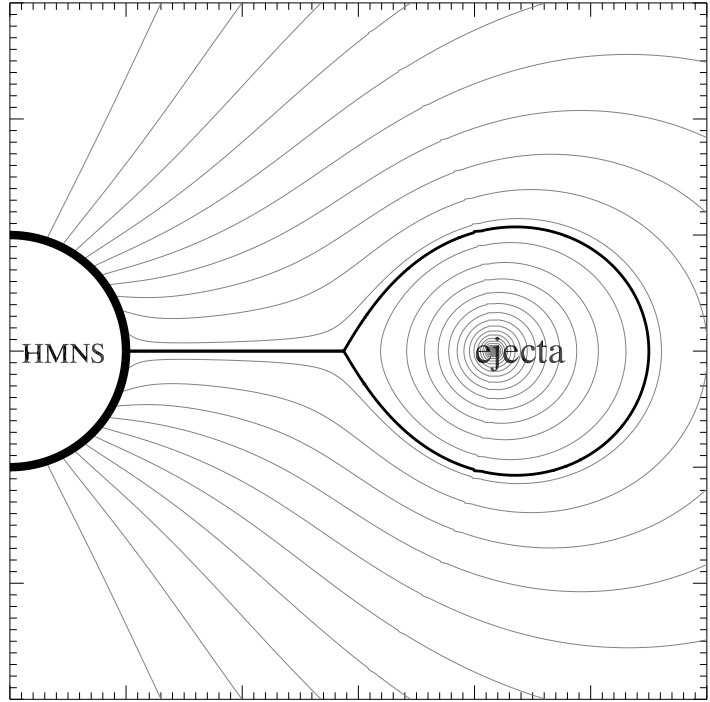}
\caption{\label{HMNS}
The illustrative scenario of our magnetic reconnection driven low luminosity GRBs. 
The thick horizontal line denotes the current sheet caused by the 
escape of the ejecta, where the Poynting flux inside the current sheet
supports the low luminosity of GRB 170817A.   
}
\end{figure}

%%%%from Yu
%We propose that the low luminosity GRB is driven by the magnetic field dissipation.
%The dynamic expansion of the ejecta drags the magnetic field lines, leading to spontaneous formation of current sheet.
Relativistic reconnection is believed to be operating in GRBs.  Magnetars, originally proposed to account for GRBs (Duncan \& Thompson 1992),  share similar behavior of low luminosity GRBs. Observations of GRB 080916C indicates that the GRB
central engine likely launches magnetically-dominated plasma and magnetic reconnection leads to the GRB prompt emissions (Zhang \& Yan 2011; McKinney \& Uzdensky 2012)  which provides a
promising mechanism to promote rapid conversion of magnetic energy to radiation.
%Relativisitc reconnection facilitate magnetic energy release and give rise to high energy emissions.
The reconnection inflow velocity is roughly $V_{\rm inflow} = M_{\rm A} V_{\rm A}$,
where $V_{\rm A}$ is the Alfv$\acute{e}$n velocity. In the magnetically dominated environment,
the Alfv$\acute{e}$n velocity is approximately the speed of light.
The magnetic reconnection model for GRBs indicates that, in the vicinity of the central engines,
the Sweet-Parker reconnection dominates and magnetic reconnection rate is rather low.
Fast reconnection switches on at a rather distant radius $\sim 10^{13} \ \rm cm$, where the ion skin 
depth becomes larger than the Sweet-Parker layer thickness, and fast Petschek-like reconnection takes place (McKinney \& Uzdensky 2012).
We also note that fast reconnection would also destroy the jet formation feasibility by the Blandford-Znajek mechanism
(Blandford \& Znajek 1977).
For these reasons, we adopt a small value of Alfv$\acute{e}$n Mach number, less than $10^{-3}$.
The toroidal electric field $E_{\phi}$ inside the current sheet is approximately
$E_{\phi} \sim B_{\rm dipole} \times (V_{\rm inflow}/c)$. 
In Figure \ref{HMNS}, we give an illustrative description of the our magnetically driven
low luminosity GRB scenario. The radiative emission comes from the Poynting flux associated with
the current sheet and the energy dissipation rate is written as $\dot{E}_{\rm diss}= c \int_{r_1}^{r_2} r E_{\phi} B_r dr$,
where $r_1$ and $r_2$ are the two tips of the current sheet. Note that both $B_{r}$ and $E_{\phi}$ vary with radius
beteween $r_1$ and $r_2$.
%The Poynting flux inside the current sheet can be
%calculated as, 
The luminosity can be roughly estimated
$M_{\rm A}  c \times (B/10^{14} \ {\rm G})^2 (r/15 \ {\rm km})^2  \sim 10^{47} \ \rm erg \ s^{-1}$.
Here the magnetic field we adopt is the average magnetic field.

The magnetic field is so strong that the dynamics is dominated by the magnetic field.
Under such circumstances, the mass of the ejecta can be estimated as
$\sim 10^{30} \times (B/10^{16} \ {\rm G})^2 (R/15 \ {\rm km})^3 \ \rm  g$,
where we assume that the ejecta is generated in the vicinity of the central neutron star
and the strength of magnetic field is adopted as $B\sim 10^{16}{\rm G}$. 
Our simulations show that the ejecta is magnetically driven and can be accelerated to the speed about $\sim 0.1c$. 
This is consistent with the observational constraints about the ejecta of the kilonova (Villar et al. 2017).
A detailed description of our model and simulation results will be reported elsewhere (Yu et al. in prep.).

%%%%

The initial pulse of GRB 170817A has peak energy about $200\ \rm keV$ (Goldstein et al. 2017).
And it is not detected in the $0.2-5 \ \rm MeV$ energy range (Li et al. 2018). The emission
following the initial pulse of GRB 170817A has even lower peak energy, about $30 \ \rm keV$
(for a thermal spectrum, Goldstein et al. 2017).
For the giant flare of magnetar SGR 1806-20, it has peak energy about $0.5-1 \ \rm MeV$ (Hurley et al. 2005; Palmer et al. 2005).
If the hard X-ray photons are due to resonant inverse Compton scattering\footnote{In the case of strong magnetic field, 
the Thomson scattering cross section is significantly reduced (Herold 1979). And the non-magnetic Thomson scattering
and inverse Compton scattering is changed to the resonant cyclotron scattering and resonant inverse Compton scattering.}
 (other radiation mechanism is also possible, Elenbaas et al. 2017),
the hard X-ray photons can only escape when the magnetic field is smaller than $10^{13} \ \rm G$,
in order to avoid photon spliting or pair production (Beloborodov 2013). The electron cyclotron energy
is about $\hbar \omega_{B} \approx 100 \ \rm keV (B/10^{13} \,\rm G)$. The resonant condition requires that
the seed photons should have typical energy $\gamma 3 kT \approx \hbar \omega_{B}$ (You et al. 2003, where $\gamma$ is the electron Lorentz factor, and assuming thermal seed photons).
The scattered photons have typical energy about $\gamma \hbar \omega_{B}$.
For a Lorentz factor of $\gamma \sim 5$, the scattered photons have an energy about $0.5\,\rm MeV$.
And the seed photons have typical energy about $20\, \rm keV$.
In the case of giant flares from magnetars, the central neutron star is always there.
And the large scale strong dipole magnetic field is always present.
This may ensure that magnetar giant flares can have a hard spectra,
especially for the initial spike (Elenbaas et al. 2017). However, in the magnetically driven origin for GRB 170817A,
the central neutron star collapses to a black hole.  During the magnetic reconnection process, the magnetic field
will decrease with time. This may explain why the initial pulse of GRB 170817A has a soft spectra compared with that of
magnetar giant flares. The softer emission following the initial pulse may be the seed thermal emission.
It may come from the photospheric emission of the fireball generated during the giant-flare-like event. Here the ``fireball''
should be similar to that generated during magnetar giant flares (Thompson \& Duncan 1995; Elenbaas et al. 2017). 
And it is different from the ``fireball'' of canonical GRBs, although the terminology ``fireball'' are used in both cases. 

During the collapse of the hypermassive neutron star to a black hole,
both matter and magnetic field can be ejected.
The ejection of magnetic field energy is also employed to explain the energy of fast radio bursts
(Falche \& Rezzolla 2014, for the collapse of a supramassive neutron star).
It can not be excluded that a fast radio burst is also generated during the collapse of a hypermassive neutron star.

A general picture for GW170817/GRB 170817A may be that: The merger of the binary neutron star may result in the birth of a hypermassive neutron star. 
After about one second, the hypermassive neutron star collapses to a black hole. During this process, a relativistic jet and a  
subrelativisitc outflow may be generated by the central engine. At the same time, the collapse of the hypermassive neutron star
will also triger the release of the magnetic energy. Since the jet is seen widely off-axis, its prompt emission may be missed. 
Instead, the magnetic energy release may be responsible for the low luminosity GRB 170817A. The subrelativistic outflow
is responsible for the kilonoa emissions. The relativistic jet may be successful or failed during its drill through the ambient 
matter. A cocoon may be generated during this process. An off-axis structured jet or near isotropic outflow (e.g., cocoon or outflow
generated by the magnetic energy release) may be responsible for the X-ray/radio afterglow (D'Avanzo et al. 2018).

\section{Discussion}

Here we provide an alternative explanation for the low luminosity GRB 170817A, associated with GW170817. Compared
with other explanations (off-axis jet, structured jet, cocoon), a magnetically driven origin naturally results in a burst
luminosity similar to that of magnetar giant flares. 
Since the giant flares of magnetars are not assumed to have a strong beaming (Lyutikov 2006),
so may be the giant-flare-like event.
Bursts with similar luminosities may also be observed in future binary neutron star merger events, provided
that they are close enough. On the other hand, if future observations found that this GRB 170817A event is singular, 
then the magnetically driven origin may be ruled out.

If we are lucky that the prompt emission of the jet is also seen, then we should see two bursts
following the gravitational wave event. The internal collision of shocks may results in a delay between the prompt emission and  magnetic energy release. This will result in a precursor (which is due to the giant-flare-like event), followed by a classical GRB. Previous observations found some possible precursors of short GRBs (Troja et al. 2010).
It is possible that these precursors are also due to magnetic energy release of the central engine.
During the merger of a neutron star/black hole system, giant-flare-like event may also happen
if strong magnetic field can also be generated (Wan 2017).

The possible contribution of the hypermassive neutron star to GRB 170817A is explored in this paper.
The magnetic energy release of the hypermassive neutron star may also contribute to the X-ray/radio afterglows
 (Salafia et al. 2017; D'Avanzo et al. 2018). The central compact object of binary neutron star merger may also be a long lived 
 neutron star, instead of a short lived hypermassive neutron star (Dai et al. 2006; Fan \& Xu 2006; Lu et al. 2015). 
 For GW170817/GRB 170817A, a long lived neutron star can not be ruled out (Ai et al. 2018) and 
 it may contribute to the kilovova emissions (Yu \& Dai 2017) . In the case of a long lived neutron star, 
 a giant-flare-like event is also possible. However, it may not correspond to the case of GRB 170817A. 
 This possibility may be revealed in future observations.

During the preparation process, we noted the paper of Salafia et al. (2017), who employed a fireball powered by a giant flare
to explain both GRB 170817A and the X-ray/radio afterglow. It is different from our scenario. In our scenario, (1) the time delay
between GRB 170817A and GW170817 is due to delayed collapse of the hypermassive neutron star, 
(2) the giant-flare is triggered by the  collapse of the hypermassive neutron star to a black hole, 
(3) we focus on the giant-flare-like origin for the low luminosity GRB 170817A.

\section*{Acknowledgments}
The authors would like to thank X.Y. Wang, Z. Li, S.L. Xiong, P.H.T. Tam, and H.G. Wang for discussions.
H.Tong is supported by NSFC (11773008).
C.Y. has been supported by NSFC (11373064,  11521303, 11733010),
%Open Research Program in Key Lab
%for the Structure and Evolution of Celestial Objects (Grant
%OP201301),
Yunnan Natural Science Foundation (2014HB048)
and Yunnan Province (2017HC018).

\label{lastpage}


\begin{thebibliography}{99}

\bibitem{Abbott2017a}
Abbott, B. P., et al., 2017a, PRL, 119, 161101

\bibitem{Abbott2017b}
Abbott, B. P., et al., 2017b, ApJL, 848, L12

\bibitem{Abbott2017c}
Abbott, B. P., et al., 2017c, ApJL, 848, L13

\bibitem{Ai2018}
Ai, S., Gao, H., Dai, Z. G., et al., 2018, arXiv:1802.00571

\bibitem{Baumgarte2000}
Baumgarte, T. W., Shapiro, S. L., Shibata, M., 2000, ApJL, 528, L29

\bibitem{Beloborodov2013}
Beloborodov, A. M., 2013, ApJ, 762, 13	

\bibitem{Blanchard2017}
Blanchard, P. K., Burger, E., Fong, W., et al., 2017, ApJL, 848, L22

\bibitem{Blandford1977}
Blandford, R. D.,  Znajek, R. L., 1977, MNRAS, 179, 433

\bibitem{Ciolfi2017}
Ciolfi, R., Kastaun, W., Giacomazzo, B., et al., 2017, PRD, 95, 063016

\bibitem{Coulter2017}
Coulter, D. A., Foley, R. J., Kilpatrick, C. D., et al., 2017, Science, 10.1126/science.aap9811 (arXiv:1710.05452)

\bibitem{Dai2006}
Dai, Z. G., Wang, X. Y., Wu, X. F., Zhang, B., 2006, Science, 311, 1127

\bibitem{DAvanzo2018}
D'Avanzo, P., Campana, S., Ghiselline, G., et al., 2018, arXiv:1801.06164

\bibitem{DT1992}
Duncan, R. C., Thompson, C., 1992, ApJ, 392, L9

\bibitem{Elenbaas2016}
Elenbaas, C., Watts, A. L., Turolla, R., et al., 2016, MNRAS, 456, 3282	

\bibitem{Elenbaas}
Elenbaas, C., Huppenkothen, D., Omand, C., et al., 2017, MNRAS, 471, 1856

\bibitem{Falcke2014}
Falcke, H.,  Rezzolla, L., 2014, A\&A, 562, A137

\bibitem{Fan2006}
Fan, Y. Z., Xu, D., 2006, MNRAS, 372, L19	

\bibitem{Fong2017}
Fong, W., Burger, E., Blanchard, P. K., et al., 2017, ApJL, 848, L23

\bibitem{Giacomazzo2013}
Giacomazzo, B.,  Perna, R., 2013, ApJL, 771, L26

\bibitem{Goldstein2017}
Goldstein, A., Veres, P., Burns, E.,  et al., 2017, ApJL, 848, L14

\bibitem{Hallinan2017}
Hallinan, G., Corsi, A., Mooley, K. P., et al., 2017, Science, 10.1126/science.aap9855 (arXiv:1710.05435)

\bibitem{Herold1979}
Herold, H., 1979, PRD, 19, 2868

\bibitem{Hurley2005}
Hurley, K., Boggs, S. E., Smith, D. M., et al, 2005, Nature, 434, 1098

\bibitem{Hotokezaka2013}
Hotokezaka, K., Kuichi, K., Kyutoku, K., et al., 2013, PRD, 88, 044026	

\bibitem{Kasliwal2017}
Kasliwal, M. M., Narkar, E., Singer, L. P., et al., 2017, Science, 10.1126/science.aap9455 (arXiv:1710.05436)

\bibitem{Li2017}
Li, T. P., Xiong, S. L., Zhang, S. N., et al, 2018, Science China Physics, Mechanics \& Astronomy, 61, 031011 (arXiv:1710.06065)

\bibitem{Lu2015}
Lu, H. J., Zhang, B., Lei, W. H., et al., 2015, ApJ, 805, 89

\bibitem{Lyne2004}
Lyne, A. G., Burgay, M., Kramer, M., et al., 2004, Science, 303, 1153

\bibitem{Lyutikov2006}
Lyutikov, M., 2006, MNRAS, 367, 1594

\bibitem{Margalit2017}
Margalit, B., Metzger, B. D., 2017, ApJL, 850, L19

\bibitem{McKinney2012}
McKinney, J. C.,  Uzdensky, D. A., 2012, MNRAS, 419, 573

\bibitem{Mereghetti2008}
Mereghetti, S., 2008, A\&ARv, 15, 225	

\bibitem{Murguia-Berthier2017}
Murguia-Berthier, A., Ramirez-Ruiz, E., Kilpatrick, C. D., et al., 2017, ApJL, 848, L34

\bibitem{Palmer2005}
Palmer, D. M., Barthelmy, S., Gehrels, N., et al., 2005, Nature, 434, 1107

\bibitem{Rezzolla2017}
Rezzolla, L., Most, E. R., Weih, L. R., 2018, ApJL, 852, L25

\bibitem{Ruiz2017}
Ruiz, M., Shapiro, S. L.,  Tsokaros, A., 2018, PRD, 97, 021501

\bibitem{Salafia2017}
Salafia, O. S., Ghisellini, G., Ghirlanda, G., et al., 2017, arXiv:1711.03112

\bibitem{Savchenko2017}
Savchenko, V., Ferrigno, C., Kuulkers, E., et al., 2017, ApJL, 848, L15

\bibitem{Sekiguchi2017}
Sekiguchi, Y., Kiuchi, K., Kyutoku, K., et al., 2011, PRL, 107, 051102

\bibitem{Shiata2017}
Shibata, M., Fujibayashi, S., Hotokezaka, K., et al., 2017, PRD, 96, 123012

\bibitem{He2017}
He, X. B., Tam, P. H. T., Shen, R. F., 2017, arXiv:1710.05869

\bibitem{Thompson1995}
Thompson, C.,  Duncan, R. C., 1995, MNRAS, 275, 255

\bibitem{Troja2010}
Troja, E., Rosswog, S.,  Gehrels, N., 2010, ApJ, 723, 1711

\bibitem{Troja2017}
 Troja, E., Piro, L., van Eerten, H., et al., 2017, Nature, 551, 71

\bibitem{Villa2017}
Villar, V. A., Guillochon, J., Burger, E., et al., 2017, ApJL, 851, L21

\bibitem{Wan2017}
Wan, M. B., 2017, PRD, 95, 104013

\bibitem{You2003}
You, J. H., Chen, W. P., Zhang, S. N., et al., 2003, MNRAS, 340, 687

\bibitem{Yu2017}
Yu, Y. W., Dai, Z. G., 2017, arXiv:1711.01898

\bibitem{Yu2013}
Yu, C.,  Huang, L., 2013, ApJL, 771, L46

\bibitem{Zhang2011}
Zhang, B.,  Yan, H., 2011, ApJ, 726, 90

\bibitem{Zrake2013}
Zrake, J.,  MacFadyen, A. I., 2013, ApJL, 769, L29

\end{thebibliography}
\end{document}